# High Albedos of Low Inclination Classical Kuiper Belt Objects

M. J. Brucker[a,b], W. M. Grundy[a], J. A. Stansberry[c], J. R. Spencer[d], S. S. Sheppard[e], E. I. Chiang[f], M. W. Buie[d,(a)]

[a]Lowell Observatory, 1400 W. Mars Hill Rd., Flagstaff, AZ 86001.
[b]Homer L. Dodge Dept. of Physics and Astronomy, University of Oklahoma, 440 W. Brooks St., Norman, OK 73019.
[c]Steward Observatory, University of Arizona, 933 N. Cherry Ave., Tucson, AZ 85721.
[d]Southwest Research Institute, 1050 Walnut St. #300, Boulder, CO 80302.
[e]Carnegie Inst. of Washington, 5241 Broad Branch Rd. NW, Washington, DC 20015.
[f]Astronomy Dept., University of California, Berkeley, 601 Campbell Hall, Berkeley, CA 94720.

**ABSTRACT**

We present observations of thermal emission from fifteen transneptunian objects (TNOs) made using the *Spitzer Space Telescope*. Thirteen of the targets are members of the Classical population: six dynamically hot Classicals, five dynamically cold Classicals, and two dynamically cold inner Classical Kuiper Belt Objects (KBOs). We fit our observations using thermal models to determine the sizes and albedos of our targets finding that the cold Classical TNOs have distinctly higher visual albedos than the hot Classicals and other TNO dynamical classes. The cold Classicals are known to be distinct from other TNOs in terms of their color distribution, size distribution, and binarity fraction. The Classical objects in our sample all have red colors yet they show a diversity of albedos which suggests that there is not a simple relationship between albedo and color. As a consequence of high albedos, the mass estimate of the cold Classical Kuiper Belt is reduced from approximately 0.01 $M_\oplus$ to approximately 0.001 $M_\oplus$. Our results also increase significantly the sample of small Classical KBOs with known albedos and sizes from 21 to 32 such objects.

## 1. Introduction

Small, undifferentiated transneptunian objects (TNOs) may be the most pristine Solar System objects available for scientific study (see reviews by Chiang et al. 2007, Morbidelli et al. 2008). Investigation of their thermal, chemical, and dynamical properties can tell us much about their history and the conditions of the early Solar System. Distinct dynamical subgroups of TNOs have been identified which are likely to have different histories. We have explored a subset of TNOs, the Classical Kuiper Belt Objects (KBOs) (TNO and KBO are sometimes used interchangeably), to determine if there exist dissimilarities in albedo with inclination within this subset.

We include in our analysis those objects which fall into the Classical regime (the main Kuiper Belt) in either of the two major TNO classification systems described by the Deep Ecliptic Survey (DES) team (Elliot et al. 2005) and described by Gladman et al. (2008). Both classification systems rely on numerical integrations over 10 Myr to determine orbital behavior. The main difference between the two classification systems lies in the definition of Scatter(ing/ed) Disk Objects (SDOs). According to Gladman et al. (2008), SDOs are only the objects "currently scattering actively off Neptune" as evidenced by a swiftly changing semi-major axis in the orbital integrations, specifically a change greater than or equal to 1.5 AU. In contrast, the DES system uses a time-averaged Tisserand parameter less than 3 relative to Neptune to separate the Scattered-Near KBOs from the Classical KBOs. The Detached TNOs of Gladman et al. (2008) have eccentricities greater than 0.24 while the DES Scattered-Extended KBOs have Tisserand parameters greater than 3 relative to Neptune and eccentricities greater than 0.2 (Elliot et al. 2005). As a result, some objects that are Scattered-Near and Scattered-Extended in the DES system are Classical KBOs according to Gladman et al. (2008).

Three subdivisions of Classical KBOs relevant to our analysis are inner Classical KBOs and dynamically hot and cold Classical KBOs. Inner Classical KBOs are Classical KBOs which have mean semi-major axes less than 39.46 AU (interior to the 3:2 resonance). Cold Classical KBOs orbit close to the invariant plane with low inclinations and eccentricities (cf. Chiang and Choi 2008). Hot Classical KBOs are dynamically excited with higher inclinations and eccentricities. We use an inclination of 5˚ to separate the hot and cold Classical KBOs for illustrative purposes as the two populations overlap.

Cold Classicals have been found to be significantly distinct from other Classicals. They are uniformly red unlike the hot Classical population which contains a continuum of colors (Tegler and Romanishin 2000, Trujillo and Brown 2002, Doressoundiram et al. 2002, Tegler et al. 2003, Gulbis et al. 2006, Chiang et al. 2007, Doressoundiram et al. 2008, Morbidelli et al. 2008). No hot Classicals are red for inclinations greater than 20˚ (Peixinho et al. 2004). A new analysis by Peixinho et al. (2008) introduces a 12˚ inclination break such that Classical KBOs with inclinations less than 12˚ relative to the Kuiper Belt Plane are red and Classical KBOs with inclinations greater than or equal to 12˚ are bluer. In addition to their different color statistics, objects with absolute visual magnitudes brighter than 6.5 tend to have higher inclinations than objects with absolute magnitudes fainter than 6.5 (Levison and Stern 2001). Therefore hot Classicals are more likely to be intrinsically brighter than cold Classicals (Morbidelli et al. 2008).

Color and magnitude are not the only properties in which dynamically cold and hot Classical KBOs differ. A much higher fraction of cold KBOs are binary than hot KBOs. Noll et al. (2008b) found that for inclinations less than 5.5˚, 17 out of 58 objects examined were binary systems (a binary fraction of $29.3^{+7.3}_{-6.4}\%$) while for inclinations greater than 5.5˚, only 4 out of 43 objects examined were binary (a binary fraction of $9.3^{+6.7}_{-4.4}\%$).

The differences in color and magnitude between hot and cold Classical KBOs have no widely accepted explanation. Though details vary substantially between proposals (e.g. Levison et al. 2008, Chiang et al. 2007), it is generally hypothesized that hot Classicals formed closer to the Sun and were scattered outward by Neptune onto highly inclined and eccentric orbits while cold Classicals formed at more distant locations and were left relatively dynamically cold. To explain the color and magnitude differences, one presumes differences in surface composition and/or physical size with heliocentric distance of formation. This presumption does not yet have quantitative support. The scattering mechanism itself for generating hot Classicals—how to raise perihelia away from Neptune so that these objects are stable over the age of the Solar System—is also unclear (e.g. Gladman and Chan 2006, Ford and Chiang 2007, Levison et al. 2008, Morbidelli et al. 2008). In particular, the scattering mechanism must have an efficiency large enough to be compatible with the small fraction of the primordial mass contained in objects with radii greater than 50 km (i.e. the sizes currently observable; Chiang et al. 2007, Kenyon and Luu 1999). This efficiency has yet to be attained via modeling.

Differences between the physical properties of the hot and cold Classical KBOs suggest that they may have formed in distinct locations within the proto-planetary disk. This case would become even stronger if a difference between the albedos of cold KBOs with respect to hot KBOs was found.

We have tested whether the dynamically cold Classical KBOs have different albedos than dynamically hot Classical KBOs, reporting new results on the albedos and radii of fifteen TNOs, thirteen of which are Classical according to Gladman et al. (2008). These results are based on *Spitzer Space Telescope* observations of thermal emissions at wavelength bands centered about 24 μm and 70 μm. Details of these observations are presented in § 2. § 3 describes how we chose which thermal model was most appropriate for our analysis. The reduction and analysis of the thermal emissions are described in § 4. The results of our analysis are presented in § 5 followed by a discussion and conclusions in §§ 6 and 7 respectively.

## 2. Observations

Fifteen TNOs were observed with the Multiband Imaging Photometer for Spitzer (MIPS) 24 μm and 70 μm channels of the *Spitzer Space Telescope* (*SST*) (Rieke et al. 2004) as part of program ID 3542 during cycle 1. The observation targets include thirteen Classical KBOs which increases the number of small Classical KBOs imaged by *SST* from 21 to 32. In our sample, two are inner Classicals, six are dynamically hot, and five are dynamically cold (Table 1). We have adopted 5˚ as the dividing inclination between hot and cold Classicals for illustrative purposes. The remaining two observation targets are classified as a Neptune Trojan and a 3:2 Resonant TNO. Of the six hot Classical KBOs, Quaoar and Altjira are known binaries (Noll et al. 2008a). 2001 $RZ_{143}$, a cold Classical KBO, is also a binary (Noll et al. 2008a). 2002 $KW_{14}$, a hot Classical KBO, is identified as a Scattered-Extended TNO in the DES classification scheme.

Table 1
Orbital Properties of *Spitzer Space Telescope* Cycle 1 Program 3542 Targets

| Provisional Designation[a] | a (AU)[b] | e[b] | i (°)[b,c] | $H_V$[d] | V-R | DES Type[e] | Name and Number |
|---|---|---|---|---|---|---|---|
| Hot Classicals | | | | | | | |
| 2001 $KA_{77}$ | 47.4830 | 0.0944 | 13.2174 | 5.608 ± 0.042[f,g] | 0.695 ± 0.081[f,g,h] | C | |
| 2002 $GJ_{32}$ | 44.2771 | 0.1061 | 13.1642 | 6.12 ± 0.13[g] | 0.68 ± 0.08[g] | C | 182934 |
| 1996 $TS_{66}$ | 43.9628 | 0.1298 | 8.8280 | 6.48 ± 0.16[i] | 0.690 ± 0.048[j] | C | |
| 2002 $LM_{60}$ | 43.3887 | 0.0417 | 8.5214 | 2.739 ±0.011[k,l] | 0.646 ± 0.012[k,l] | C | Quaoar 50000 |
| 2002 $KW_{14}$ | 46.7707 | 0.2016 | 8.4874 | 5.959 ± 0.060[m] | 0.72 ± 0.07[m] | SE | |
| 2001 $UQ_{18}$ | 44.3018 | 0.0581 | 5.4610 | 6.55 ± 0.21[g] | 0.735 ±0.078[g] | C | Altjira 148780 |
| Cold Classicals | | | | | | | |
| 2000 $OK_{67}$ | 46.5638 | 0.1401 | 4.9999 | 6.50 ± 0.10[f,n,o] | 0.653 ± 0.048[f,n,o] | C | 138537 |
| 2002 $VT_{130}$ | 42.4930 | 0.0334 | 2.7921 | 5.95 ± 0.5 | -- | C | |
| 2001 $QD_{298}$ | 42.6911 | 0.0449 | 2.7769 | 6.68 ± 0.08[g] | 0.67 ± 0.09[g] | C | |
| 2001 $RZ_{143}$ | 44.0958 | 0.0687 | 2.4295 | 6.23 ± 0.5 | -- | C | |
| 2001 $QS_{322}$ | 44.0890 | 0.0398 | 1.8803 | 6.22 ± 0.5 | -- | C | |
| Inner Classicals | | | | | | | |
| 2001 $QT_{322}$ | 37.1091 | 0.0319 | 2.5185 | 6.4 ± 0.5 | 0.53 ±0.12[p] | C | 135182 |
| 2002 $KX_{14}$ | 38.8406 | 0.0469 | 2.8165 | 4.89 ±0.03[m,q] | 0.621 ± 0.022[m,q] | C | 119951 |
| Resonant KBOs | | | | | | | |
| 2001 $QR_{322}$ | 30.1106 | 0.0311 | 1.2645 | 8.11 ± 0.02[m] | 0.46 ± 0.02[m] | 1:1 | |
| 2003 $QX_{111}$ | 39.4566 | 0.1553 | 8.3295 | 6.76 ± 0.5 | -- | 3:2 | |

[a]Objects are sorted by Gladman et al. (2008) orbital type and decreasing inclination with a break at 5° to separate the overlapping hot and cold Classical KBO populations.
[b]Values averaged over 10 Myr orbital integrations.
[c]Average inclinations are with respect to the invariable plane.
[d]Where published V magnitudes were available (as indicated by notes f-q), Hv was estimated by combining V magnitudes weighted according to their error bars, assuming G=0.15 in the Bowell et al. (1989) photometric system. The larger of the formal error or the scatter of the individual measurements was taken as the uncertainty. For the remaining objects with no published photometry, Hv was taken as a straight average of the crude photometry reported with the astrometric observations used to compute each object's orbit, subject to assumed colors and G=0.15 phase behavior. According to Romanishin and Tegler (2005), this procedure tends to overestimate the absolute brightness by about 0.3 mag, so we added 0.3 to the value and used conservative half-magnitude error bars. The linear phase function method of Sheppard and Jewitt (2002) produces similar absolute magnitudes thus the method used will not change the results.
[e]DES orbital type (Elliot et al. 2005): SE - Scattered-Extended, C - Classical, #:# - Resonant.

[f]Doressoundiram et al. (2002).
[g]Doressoundiram et al. (2005).
[h]Peixinho et al. (2004).
[i]Davies et al. (2000).
[j]Jewitt and Luu (2001).
[k]Fornasier et al. (2004).
[l]Tegler et al. (2003).
[m]The objects 2002 $KW_{14}$, 2002 $KX_{14}$ and 2001 $QR_{322}$ were observed in the V and R bands by S. Sheppard with filters based on the Johnson system. 2002 $KW_{14}$ and 2002 $KX_{14}$ were observed with four 300 second images in each filter for each object at the du Pont 2.5 meter telescope on UT July 19, 2007 with the Tek5 CCD (0.259" $pixel^{-1}$). 2001 $QR_{322}$ was observed with four 350 second images in each filter on the Magellan 6.5 meter Clay telescope on UT Nov. 3, 2005 with the LDSS3 CCD (0.189" $pixel^{-1}$).
[n]Delsanti et al. (2001).
[o]Stephens et al. (2003).
[p]Stephen Tegler, personal communication 2007.
[q]Romanishin et al. (2008b).

At least four astronomical observation request (AOR) visits were made to each object (Table 2). Each AOR consisted of many short exposures or data collection events (DCEs) executed with dithering of the field of view (FOV). The DCEs were each 10 MIPS seconds long (about 10.49 s). The AORs were timed such that the target object had moved more than the width of a point-spread function (PSF) but less than the width of the FOV in between each visit in order to improve background subtraction. With this method, unlike shadow observations, no empty fields are observed. The position of the target object on the mosaic image was also dithered among the AORs in later observations. We used the MIPS instrument team data analysis tools (Gordon et al. 2005) and other off-line tools to process the data, removing instrumental artifacts on a DCE-by-DCE basis, and produced a final calibrated mosaic for each wavelength band for each visit. Details of this processing can be found in Stansberry et al. (2008).

## 3. Thermal Modeling

Infrared emissions of TNOs can be modeled either in a complex or simple manner depending upon how many assumptions are made about their physical and chemical properties. Three common models are the Thermophysical Model (TPM), the Standard Thermal Model (STM), and the Isothermal Latitude Model (ILM).

The Thermophysical Model, based on the work of Spencer (1987), is by far the most physically realistic of the three models. The TPM uses bulk thermodynamic properties and conduction in one dimension beneath an object's surface to derive surface temperatures. Free parameters such as the heat capacity, thermal conductivity, pole position, rotation period, and thermal inertia are constant with respect to time and temperature.

The Standard Thermal Model (Lebofsky and Spencer 1989) calculates the flux emitted from a spherical object assuming that it is at instantaneous equilibrium with the incident sunlight. The STM is sometimes referred to as the slow rotator STM since a smooth body would behave in this manner in the limit of slow rotation (or zero thermal inertia).

Table 2

Observational Circumstances for *Spitzer Space Telescope* Cycle 1 Program 3542 Targets

| Provisional Designation or Name | MJD[a] | RA[a] (h) | Dec[a] (˚) | r[a,b] (AU) | Δ[a,b] (AU) | α[a] (˚) | 24 μm: #AOR[c] | 24 μm: t[d] (s) | 70 μm: #AOR | 70 μm: t[d] (s) | First AORkey |
|---|---|---|---|---|---|---|---|---|---|---|---|
| Hot Classicals | | | | | | | | | | | |
| 2001 $KA_{77}$ | 53829.1 | 16.9 | -19.6 | 48.6 | 48.4 | 1.2 | 11 | 16466 | 6 | 6105 | 15466752 |
| 2002 $GJ_{32}$ | 53785.5 | 14.7 | -20.2 | 43.2 | 43.2 | 1.3 | 6 | 7484 | 5 | 5088 | 15481856 |
| 1996 $TS_{66}$ | 53399.8 | 2.9 | 23.0 | 38.5 | 38.2 | 1.4 | 6 | 5988 | 6 | 10133 | 11096320 |
| Quaoar | 53829.0 | 17.1 | -15.4 | 43.3 | 43.1 | 1.3 | 12 | 2994 | 12 | 2140 | 15475968 |
| 2002 $KW_{14}$ | 53611.8 | 15.5 | -18.6 | 40.0 | 40.0 | 1.5 | 9 | 3368 | 5 | 5088 | 15472128 |
| Altjira | 53785.9 | 3.6 | 23.6 | 45.3 | 45.1 | 1.2 | 11 | 16466 | 6 | 6105 | 15465472 |
| Cold Classicals | | | | | | | | | | | |
| 2000 $OK_{67}$ | 53316.7 | 22.5 | -11.8 | 40.6 | 40.1 | 1.3 | 4 | 3493 | 4 | 5413 | 11112192 |
| 2002 $VT_{130}$ | 53786.1 | 4.0 | 21.9 | 42.8 | 42.4 | 1.3 | 8 | 11975 | 6 | 6105 | 15466240 |
| 2001 $QD_{298}$ | 53315.1 | 21.8 | -18.5 | 41.2 | 40.9 | 1.4 | 4 | 3992 | 4 | 6084 | 11104000 |
| 2001 $RZ_{143}$ | 53368.4 | 1.3 | 8.9 | 41.4 | 41.0 | 1.3 | 4 | 3992 | 4 | 5413 | 11107072 |
| 2001 $QS_{322}$ | 53708.0 | 23.6 | -2.9 | 42.3 | 41.9 | 1.2 | 7 | 8732 | 5 | 5088 | 15465984 |
| Inner Classicals | | | | | | | | | | | |
| 2001 $QT_{322}$ | 53366.5 | 23.6 | -0.8 | 36.9 | 36.9 | 1.6 | 6 | 3742 | 6 | 10133 | 11093248 |
| 2002 $KX_{14}$ | 53612.0 | 15.7 | -20.1 | 39.6 | 39.5 | 1.5 | 8 | 2994 | 5 | 2150 | 15463168 |
| Resonant KBOs | | | | | | | | | | | |
| 2001 $QR_{322}$ | 53368.3 | 0.2 | 0.4 | 29.7 | 29.6 | 1.9 | 4 | 998 | 4 | 6756 | 11090176 |
| 2003 $QX_{111}$ | 53366.3 | 23.9 | -4.6 | 39.5 | 39.5 | 1.5 | 5 | 4366 | 6 | 10133 | 11099392 |

[a]Values averaged over all AORs (astronomical observation requests) used.
[b]r and Δ are the distances between the Sun and the object and the Earth and the object respectively.
[c]Rejected AORkeys (designation number for an AOR) for 24 μm:
2002 $GJ_{32}$: 15482112 (A), 15482624 (A), 15482880 (A);
2002 $VT_{130}$: 15472640 (B), 15473408 (B), 15473920 (B);
2001 $QS_{322}$: 15468288 (A), 15469312 (A), 15473152 (A);
2002 $KX_{14}$: 15464960 (A);
2003 $QX_{111}$: 11099648 (A) where A is a rejection due to an asteroid and B is a rejection due to background noise.
[d]Total integrated exposure time.

The Isothermal Latitude Model (Lebofsky and Spencer 1989) is a variation on the STM in which the object is not in instantaneous equilibrium. The temperature is dependent on latitude rather than local time of day. The ILM is sometimes referred to as the fast rotator STM since a rapidly rotating body (or a body with infinitely high surface thermal inertia) would respond to sunlight in this fashion. TNOs are unlikely to have their rotational axes aligned perpendicular or parallel to the line of sight. The average of expected viewing angles is a subsolar latitude of 30˚ (Sheppard et al. 2008). The ILM can adjust the axial tilt to a desired angle with respect to the Sun-object-observer plane.

We used a Monte Carlo approach to assess uncertainties (described in § 4) which repeatedly calls a thermal model function to evaluate our observations. This method of analysis necessitated using the thermal model with the shortest computer run time. The TPM could not be used in our simulations because it has a long run time and requires knowledge of the bulk thermodynamic properties of TNOs which is lacking. Therefore, we needed to use one of the much faster STM or ILM. Both the STM and the ILM were compared to the TPM to determine which model more closely recovered the radius and albedo inputs of the TPM when their fluxes were set equal to the derived TPM fluxes (simulating TNO thermal emissions). We tested the ILM at subsolar latitudes of 0˚, 30˚, and 90˚. The ILM at 90˚ was tested to verify that its results concur with the results for the STM.

Both the STM and the ILM include a beaming factor, η, which ostensibly represents an increase in thermal emission towards the Sun but effectively adjusts the resultant flux values to compensate for qualities such as rough surfaces, rotation, thermal conduction, and correct phase integral that have not been included in the models to maintain their simplicity.

We have two main assumptions for the thermal models presented regarding the phase angle and the phase integral. First, objects are assumed to be observed at a phase angle of zero. This is permissible given that thermal emissions do not produce narrow opposition surges and that all fifteen objects in program 3542 were observed at phase angles between 1˚ and 2˚ (tolerably close to 0˚). The second assumption is that the phase integral, *q*, is a linear function of the visual geometric albedo:

$$q = 0.336 * p_V + 0.479 \qquad (1)$$

This relationship is based on published visual data for icy satellites of gas giants as no direct data are available for TNOs (Fig. 1). We have adopted the linear fit that does not include the outlying data of Europa and Phoebe; however, the resultant difference in the best fitting radius between the two phase integral functions is negligible. It should be noted that all of the satellites used to make the linear fit have reported phase integrals greater than or equal to 0.5 where 0.39 is the standard value of the phase integral assumed for dark asteroids (Bowell et al. 1989) and for TNOs in some other studies (e.g. Stansberry et al. 2008). The resultant difference in radius is less than 0.01% when the adopted phase integral function is used as opposed to a phase integral of 0.39 when infinite signal-to-noise ratios (SNRs) are assumed. The Bond albedo differs greatly with the linear function of *q* as compared to the constant value of *q* for high geometric albedos. Recall that $A=p_V*q$ where *A* is the Bond albedo. This relationship between the geometric and bolometric Bond albedos is significant when performing thermophysical modeling. An object's thermal flux depends upon its radius and Bond albedo. If one assumes that an object is spherical and that its surface is in instantaneous equilibrium, then the solar flux absorbed by the object can be equated to the thermal flux emitted by the object. From Lebofsky and Spencer (1989),

$$\pi R^2 (1 - A) S = \eta \varepsilon \sigma R^2 \int_{-\pi}^{\pi} \int_{-\pi/2}^{\pi/2} T^4(\theta,\phi) \cos\phi \, d\phi \, d\theta \qquad (2)$$

where *R* is the radius, *A* is the Bond albedo, *S* is the insolation, η is beaming factor, ε is the emissivity, σ is the Stefan-Boltzmann constant, and $T(\theta,\phi)$ is the temperature at longitude θ and latitude ϕ. Thus the temperature profile used in the thermophysical model is a function of the Bond albedo. If the geometric albedo is used to calculate the Bond albedo using an inappropriate value of the phase integral, then the temperature profile will not be as accurate as it could be.

A series of trials was conducted to determine which of the simpler models (the STM, the ILM at a subsolar latitude of 0˚, and the ILM at a subsolar latitude of 30˚) produces results the most similar to those of the TPM. Sets of thermodynamic properties (Table 3) encompassing plausible ranges of thermal parameters were input into the TPM. Along with these input parameters, an adjusted beaming factor η was calculated as a substitution for a rough surface correction at much less computational cost. We determined the beaming factor adjustment according to Spencer (1990) who observed that the behavior of his rough-surface TPM could be reasonably well approximated by the STM given a suitable value of η.

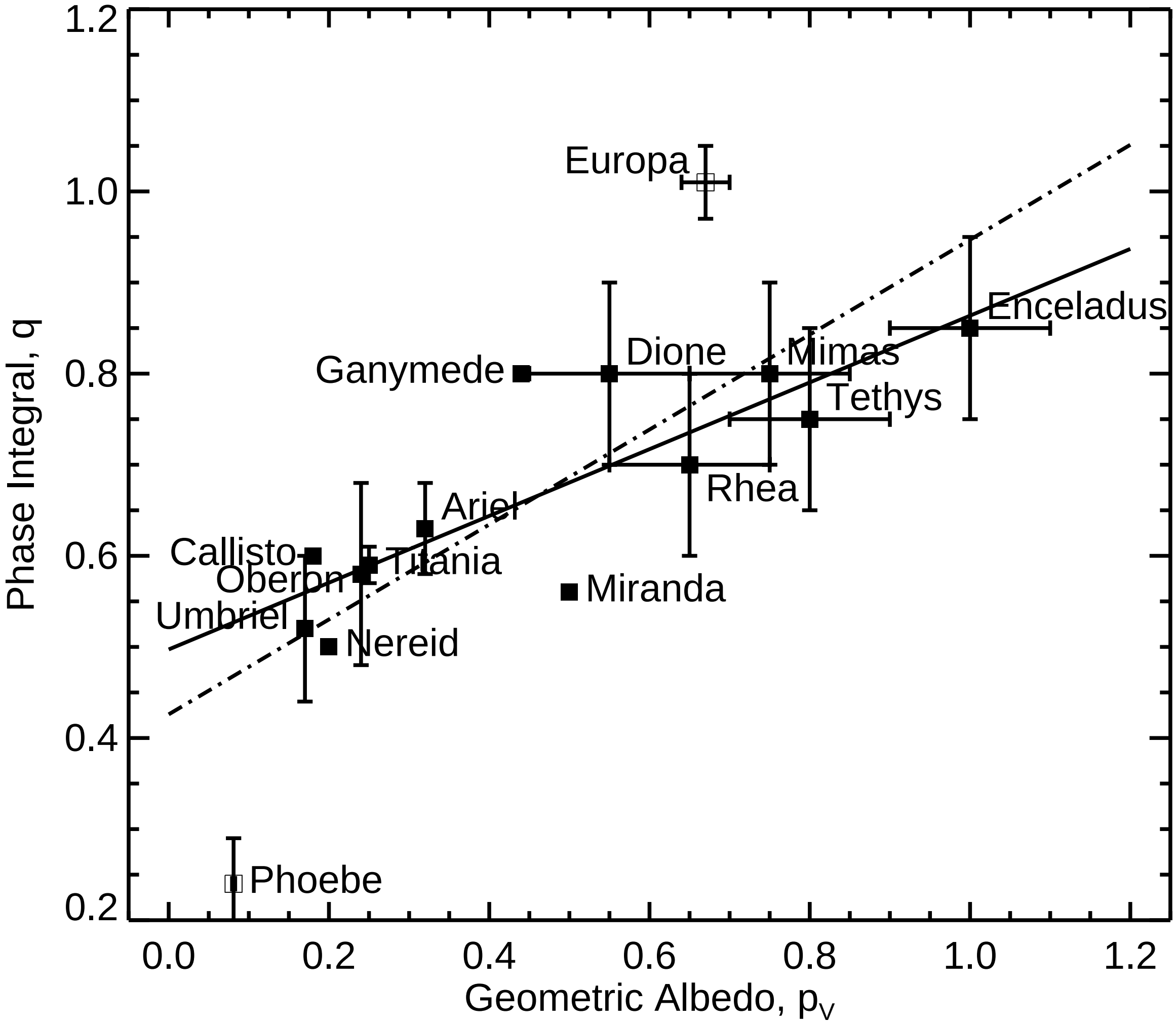

Figure 1. The visual geometric albedo, $p_V$, is plotted versus the phase integral, $q$, for icy satellites of gas giants. The solid line is the adopted best linear fit to the satellite data without including Phoebe and Europa: $q=0.336*p_V+0.497$. The dot-dashed line is the best linear fit to the data including Phoebe and Europa: $q=0.521*p_V+0.426$. Since not all of the data points had published error bars, the points were not weighted according to their errors when determining the linear fit. The data come from Buratti et al. (1990), Cruikshank and Brown (1986), Grundy et al. (2007), Morrison et al. (1986), Simonelli et al. (1999), Thomas et al. (1996), and Veverka et al. (1986).

The STM, the ILM at 0˚, and the ILM at 30˚ with three free parameters, radius, albedo, and beaming factor, were fit to the TPM fluxes. We found that both the STM and the ILM at 30˚ do reasonably well at recovering the radius and albedo from the more complex TPM under TNO-like conditions at MIPS wavelengths. The STM has an average difference in radius of -0.4% and an average difference in albedo of +0.5% when compared to the TPM in the cases tested. The ILM at 30˚ has an average difference in radius of +1.6% and an average difference in albedo of -3.6%. The ILM at 0˚ has an average difference in radius of +4.5% and an average difference in albedo of -10.0%. We chose to use the STM in our Monte Carlo simulations because it approximates the TPM well, is simpler, and is more widely used (e.g. Stansberry et al. 2008) than the ILM at 30˚.

Table 3
Thermal Model Comparison Variables and Quantities

| Quantity | Symbol | Units | Values |
|---|---|---|---|
| Subsolar Latitude | $\theta$ | ° | 0, 15, 30, 45, 60, 75, 90 |
| Bond Albedo | $A_B$ | -- | 0.01, 0.1, 0.5 |
| Wavelength | $\lambda$ | μm | 23.68, 71.42 |
| Thermal conductivity[a] | $\kappa$ | erg cm$^{-1}$ s$^{-1}$ K$^{-1}$ | 225, 111112 |
| Solar distance | r | AU | 30, 50 |
| Observer distance | $\Delta$ | AU | 29, 49 |
| Rotational Period | P | h | 5, 15 |
| Radius | R | km | 100 |
| Heat capacity[a] | c | erg g$^{-1}$ K$^{-1}$ | $10^5$ |
| Density | $\rho$ | g cm$^{-3}$ | 0.9 |
| Emissivity | e | -- | 0.9 |

[a]These values were chosen such that the thermal inertia, $\Gamma$, is 4500 and $10^5$ erg cm$^{-2}$ s$^{-1/2}$ K$^{-1}$.

## 4. Analysis

The mosaic images comprising the observations for each object were reduced and combined to produce a composite image in each wavelength band from which the object's flux was measured using PSF fitting photometry. First, the images were aligned sidereally. Then, we subtracted a constant value from each image so that all of the images would have the same median pixel value (Figs. 2a and 3a). A super-sky image was created by averaging the sidereally-aligned images together pixel by pixel while the target area was masked (Figs. 2b and 3b). If there were six or more overlapping pixels, a robust mean (cf. Buie and Bus 1992) was used instead of a standard mean. The super-sky image was subtracted from the mosaic images. If an asteroid passed through the TNO target pathway in a super-sky-subtracted image, then the image containing the asteroid was rejected, the super-sky image was recreated using the remaining images, and the new super-sky image was subtracted from the mosaic images.

The super-sky-subtracted images were aligned on the target object and averaged together pixel by pixel using the robust mean if there were six or more overlapping pixels. If the object was visible in this stacked image, then the target marker was realigned on the centroid of the object rather than the position given by the object's ephemeris.

PSF fitting photometry was used to determine flux measurements and their uncertainties. A normalized wavelength-specific and temperature-appropriate PSF provided by the Spitzer team (Engelbracht et al. 2007, Gordon et al. 2007) was fit to the target. By comparing the target region and the PSF, the free parameters (the brightness of the PSF and background brightness level) with the lowest residual $\chi^2$ were determined. The pixel brightness of the PSF that minimized $\chi^2$ was converted into the flux measurement using the 24 μm calibration factor of $1.067 \times 10^{-3}$ mJy arcsec$^{-2}$ DN$^{-1}$ (0.0454 MJy sr$^{-1}$; Engelbracht et al. 2007) or the 70 μm calibration factor of 16.5 mJy arcsec$^{-2}$ DN$^{-1}$ (702 MJy sr$^{-1}$; Gordon et al. 2007).

We estimated the uncertainty of our flux measurements by fitting the same PSF to a series of locations surrounding the target and illustrated by the circles in Figs. 2c and 3c (24 locations for 24 μm images and 20 locations for 70 μm images). The PSF background brightness level was held constant. The brightness of the PSF required to minimize $\chi^2$ for each of the locations was determined. The RMS scatter of the fitted background fluxes was taken as the flux uncertainty.

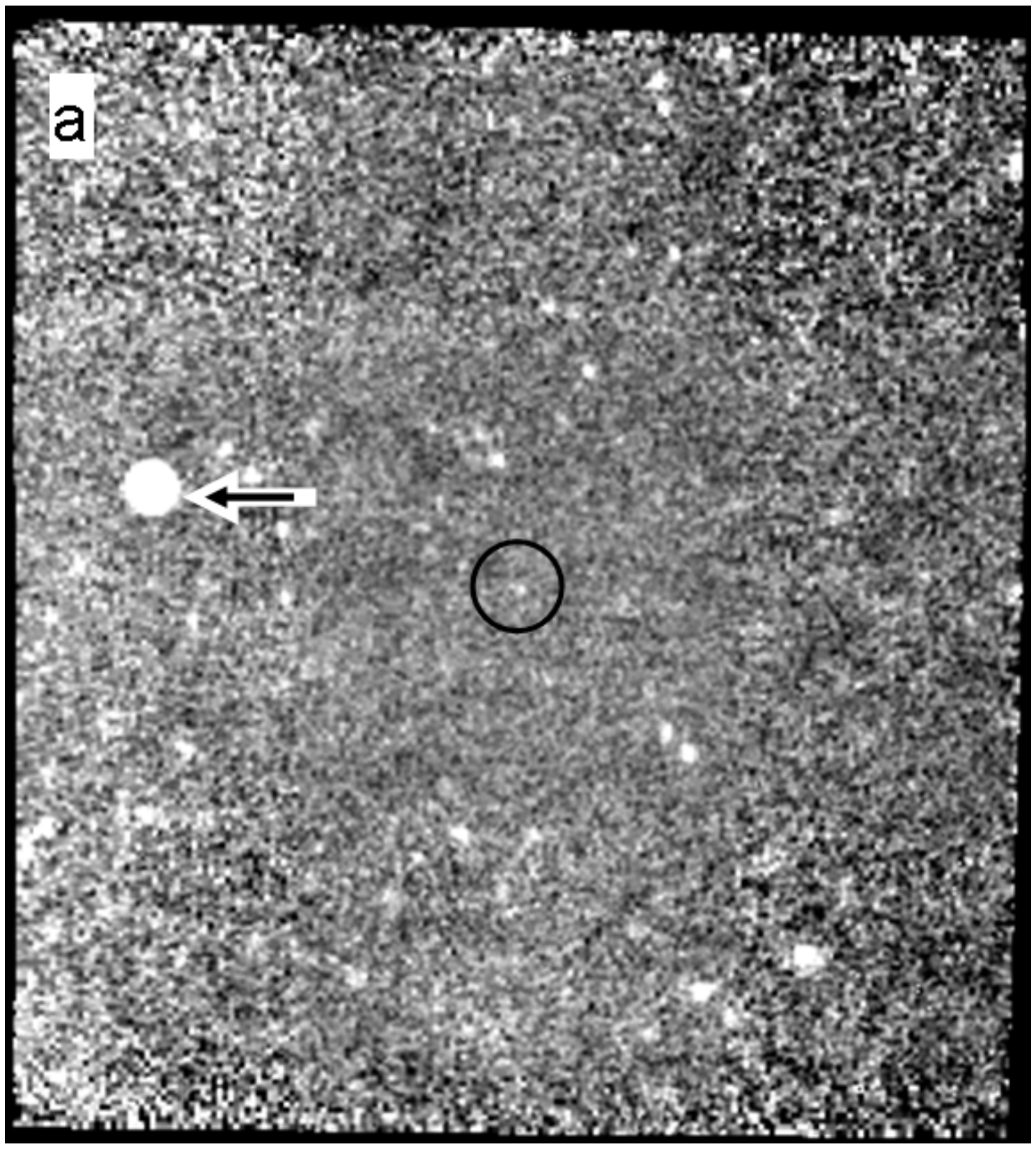

Figure 2a.

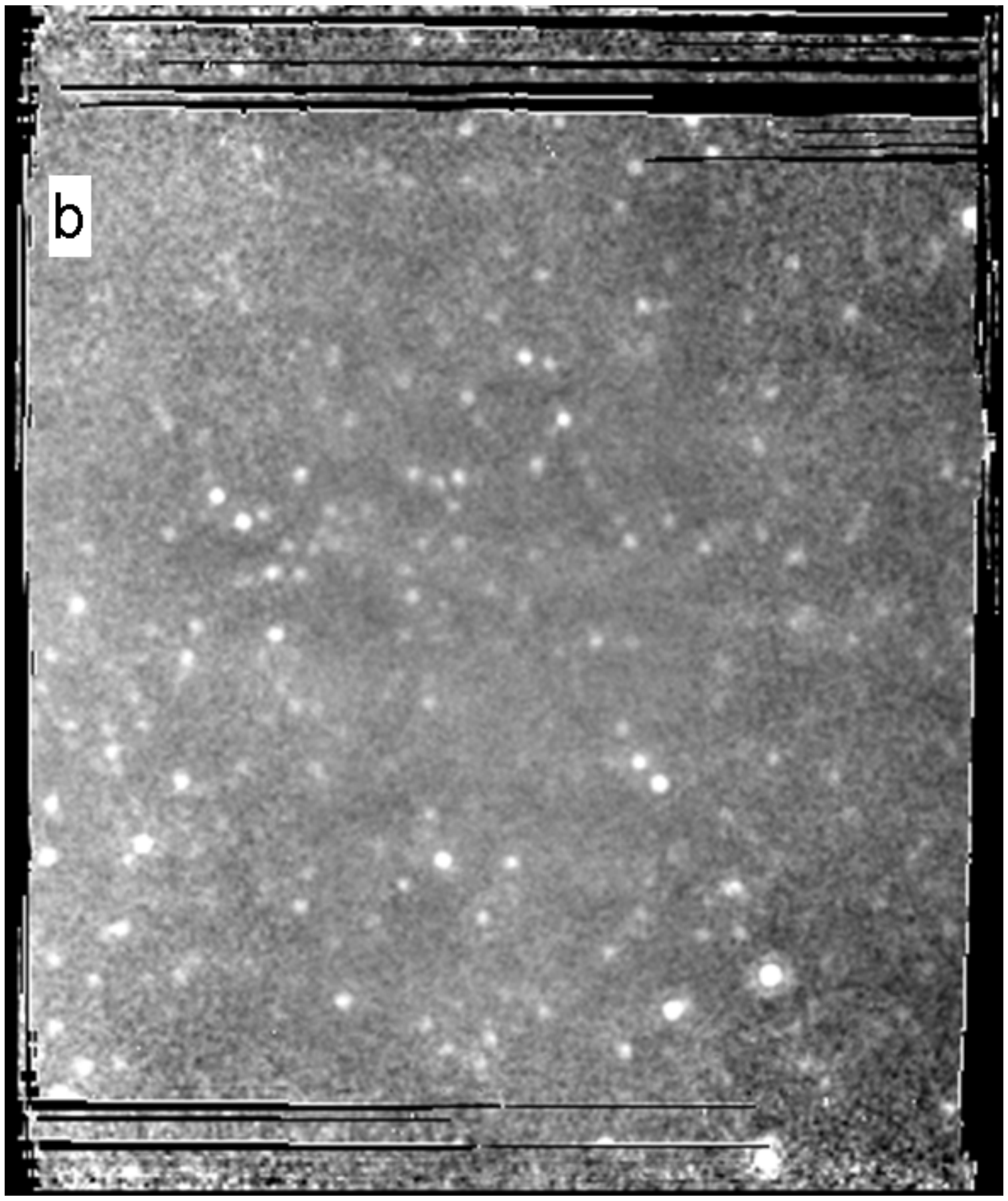

Figure 2b.

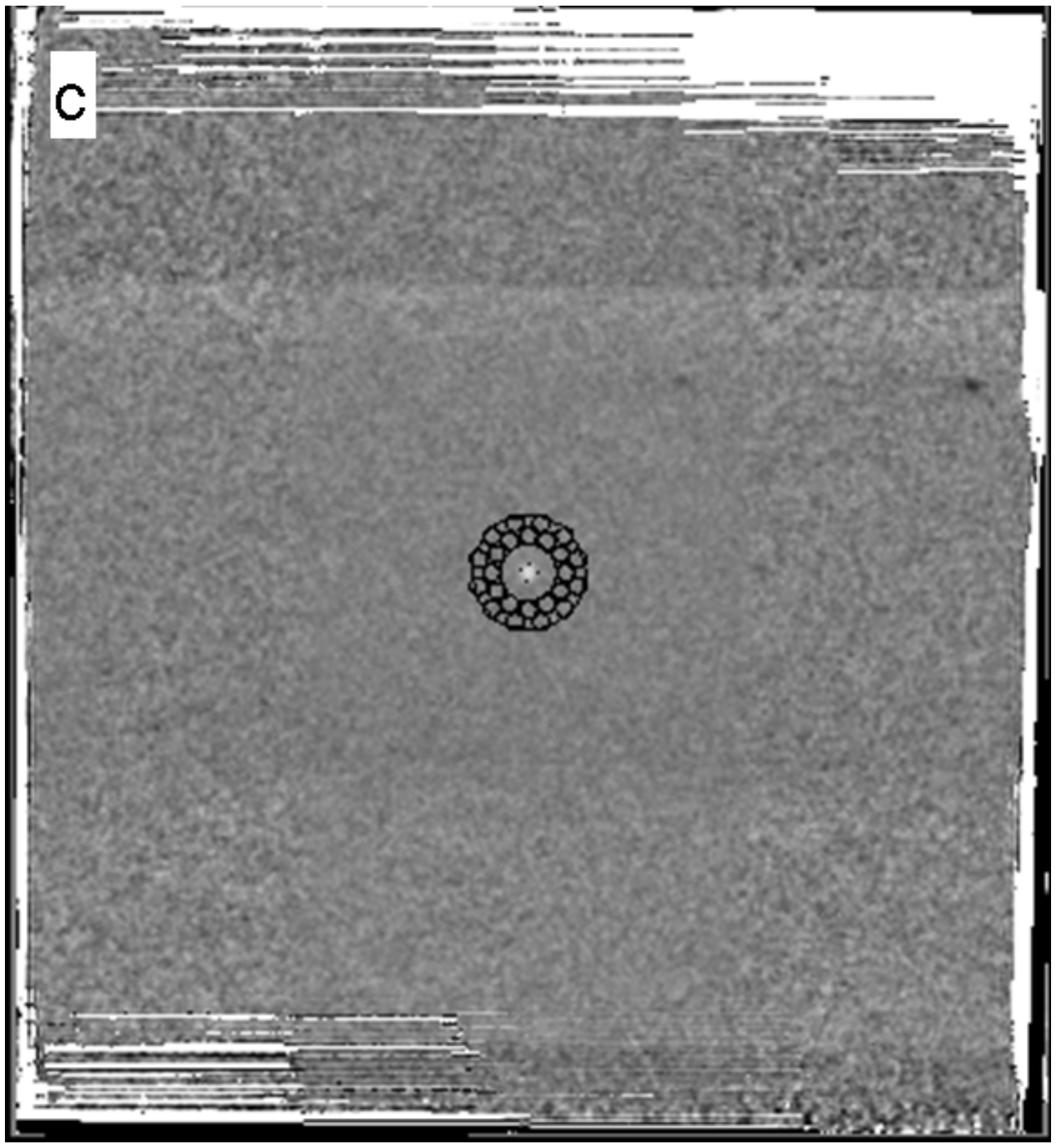

Figure 2. (a) A single mosaic image of Quaoar at 24 μm. The circle marks the position of Quaoar and the arrow points to a foreground asteroid passing through the FOV. Black lines around the edges are undefined pixels. (b) This super-sky image is the average of twelve mosaic images to facilitate background subtraction. It is larger than (a) since the composite images have slightly different FOVs due to dithering. (c) The final image is an average of twelve sky-subtracted mosaic images. 24 circles in a double ring around Quaoar mark the centers of regions used to measure the background noise. The scale is 1.245 arcsec pixel$^{-1}$ giving (c) a FOV of 7.57'x8.36'.

The flux measurements and measurement uncertainties from the 24 μm and 70 μm channel images were converted to monochromatic flux densities at 23.68 μm and 71.42 μm by applying temperature-dependent color corrections. The color corrections were found by linearly interpolating appropriate values from the table provided in Stansberry et al. (2007). For Quaoar and 2001 $QR_{322}$, which have SNRs greater than five in both wavelengths, the temperature and color corrections were computed iteratively until their individual temperatures converged on a single value. Approximate temperatures and color corrections for the other thirteen objects in our sample were interpolated using the results from Quaoar and 2001 $QR_{322}$ assuming that temperatures are proportional to $r^{-1/2}$ where $r$ is the distance from the Sun to the object (Table 4). These temperatures are only for calculating the color correction and should not be used for thermal modeling.

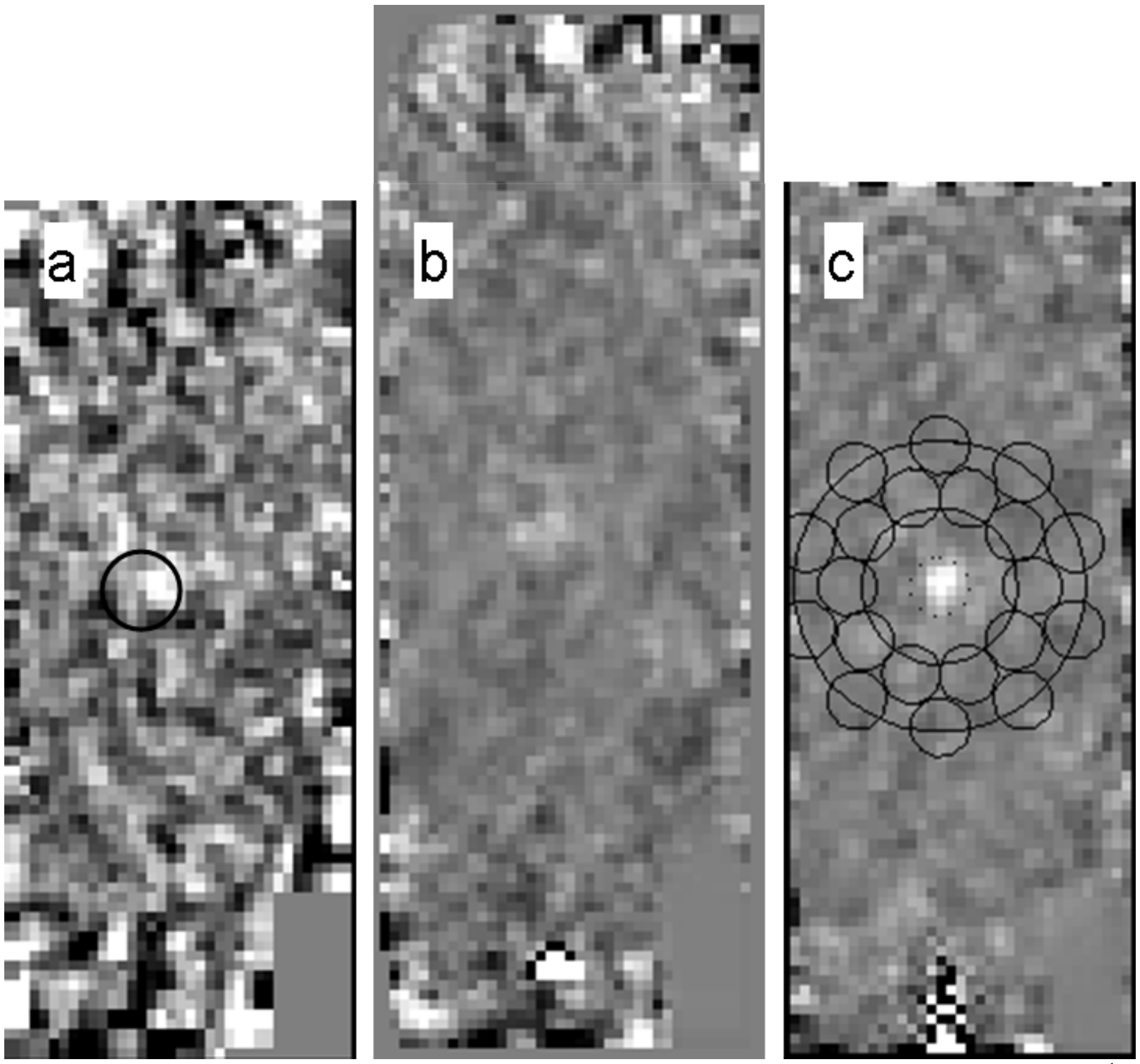

Figure 3. Similar to Figure 2 for Quaoar at 70 μm with a scale of 4.925 arcsec pixel$^{-1}$ giving (c) a FOV of 2.87'x7.39'. 20 circles mark the measured background noise regions. The 70 μm images have half the FOV of the 24 μm images due to an external cable failure that disabled half of the array and the gray box in the lower right corner of each image contains no data due to an additional cable failure (Rieke et al. 2004, Gordon et al. 2005).

The total uncertainty of our flux measurements reflects the uncertainty in our measurements and the calibration uncertainties of the MIPS channels. The uncertainty in the absolute calibration of 24 μm measurements is 2% and of 70 μm measurements is 5% (Engelbracht et al. 2007, Gordon et al. 2007). The color correction contribution to the calibration uncertainty is not likely to be significant (Stansberry et al. 2007). We have adopted values of 4% and 8% for the 24 μm and 70 μm calibration uncertainties respectively due to flux levels below the range used in calibration, extra processing steps beyond those used in calibration, and colder sources than those used in calibration. To be conservative, our total uncertainties for 24 μm flux measurements are taken to be the larger of $\sqrt{(\sigma_f)^2 + (0.04 * f)^2}$ and $1.04*(\sigma_f)$ where $\sigma_f$ is the $1\sigma$ uncertainty of the flux measurement and $f$ is the flux measurement. The former calculation is more appropriate for firm detections and the latter calculation is more appropriate for marginal detections where $f$ may be zero. The 70 μm flux uncertainties were calculated similarly.

Preliminary values for an object's radius and geometric albedo were estimated by applying the Monte Carlo method with input parameters including the object's color-corrected flux measurements, total uncertainties, absolute visual magnitude, and solar and observational distances. We chose 10,000 alternate flux and magnitude values from a normal distribution centered about the real measurements and consistent with the total errors. The Standard Thermal Model (STM) was fit to the alternate fluxes and magnitudes to determine which radius, geometric albedo (and corresponding phase integral), and beaming factor ($\eta$) produced matching flux values. The median radius, albedo, $\eta$ and their $1\sigma$ uncertainties were evaluated from that set of 10,000 results. These medians and standard deviations comprise the final results for firm detections.

Marginal detections, when either the 24 μm measurement or the 70 μm measurement had a SNR less than four and greater than one, required correction factors to the median radius, albedo, and $\eta$. The correction factors reduce the bias created by discarding randomly-generated flux values which happen to be pathological (e.g. negative thermal fluxes). They were determined via an additional layer of Monte Carlo simulations. The STM was fit to fabricated sets of 'real' flux and magnitude values in order to determine 'real' radii, albedos, and $\eta$s. Sets of 'measured' flux and magnitude values were constructed based on each set of 'real' values. Then the STM was fit to the fabricated 'measured' values to derive 'modeled' median radii, albedos, and $\eta$s. The multiplicative radial correction factor is the median of the ratios of 'real' radii to 'modeled' radii and similarly for the correction factors to the preliminary albedo and $\eta$. The correction factors are listed in Table 5.

For non-detections (i.e. SNR less than one), the correction factor to the Monte Carlo simulation is insufficient. Lower limits on albedos and upper limits on radii were determined for objects that were not detected in one or both wavelengths. In order to calculate these limits, a series of radii and corresponding albedos consistent with the absolute magnitude were input into the STM with constant $\eta$. $\eta$ was fixed at 1.94 which is the upper $1\sigma$ value of the set consisting of the Classical KBO $\eta$ values directly from Stansberry et al. (2008) and our detected Classical KBO $\eta$ values (not including any inner Classical KBOs). The radius at which the STM output flux exceeds the upper $1\sigma$ detection level of the observed flux is reported as the upper limit on the radius.

## 5. Results

We present the albedos and radii of fifteen TNOs derived from thermal emissions. All of the inner and cold Classical KBOs in our sample have high albedos. The hot Classical KBOs show a more diverse mix of albedos. The fluxes, uncertainties, temperatures, along with derived albedos, radii, and beaming factors are given in Table 4. Four of the six hot Classical KBOs have albedos less than 0.18. The fifth was detected at 70 μm but not at 24 μm and may have a low albedo. The sixth hot Classical KBO was not detected at 70 μm and may have a high albedo; however, its mean inclination is 5.46° which lies in the transitional area in which an object could belong to either sub-population (it is important to remember that the hot and cold Classical populations overlap in orbital element space). The mean, median, and standard deviation of the albedos are 0.11, 0.12, and 0.06 respectively for the four detected hot Classical KBOs. One of the five cold Classical KBOs has a moderate albedo of $0.18\pm^{0.17}_{0.08}$ and the remaining four cold Classical KBOs were non-detections. One of the inner Classical KBOs has a high albedo of $0.60\pm^{0.36}_{0.23}$ and the other was not detected at 70 μm. The Neptune Trojan and the 3:2 Resonant KBO have low albedos.

Table 4
Results for *Spitzer Space Telescope* Cycle 1 Program 3542 Targets

| Designation | 24 μm: Flux[a] (mJy) | 70 μm: Flux[a] (mJy) | $T^b$ (K) | $p_V{}^c$ | Radius[c] (km) | $\eta^c$ |
|---|---|---|---|---|---|---|
| Hot Classicals | | | | | | |
| 2001 $KA_{77}$ | 0.0077 ± 0.0023 | 4.12 ± 0.77 | 47 | $0.0250\pm^{0.0095}_{0.0080}$ | $317\pm^{67}_{46}$ | $2.80\pm^{0.51}_{0.39}$ |
| 2002 $GJ_{32}$ | 0.0101 ± 0.0060 | 1.45 ± 0.86 | 51 | $0.12\pm^{0.14}_{0.06}$ | $112\pm^{44}_{35}$ | $1.78\pm^{0.74}_{0.60}$ |
| 1996 $TS_{66}$ | 0.0935 ± 0.0048 | 2.99 ± 0.90 | 55 | $0.120\pm^{0.072}_{0.047}$ | $97\pm^{27}_{19}$ | $0.96\pm^{0.27}_{0.18}$ |
| Quaoar[d,e] | 0.2241 ± 0.0056 | 24.91 ± 2.14 | 51 | $0.172\pm^{0.055}_{0.036}$ | $454\pm^{56}_{59}$ | $1.51\pm^{0.18}_{0.20}$ |
| 2002 $KW_{14}$ | < 0.0060 | 3.30 ± 1.09 | 54 | > 0.05 | < 180 | -- |
| Altjira[d] | 0.0167 ± 0.0025 | < 0.85 | 49 | > 0.10 | < 100 | -- |
| Cold Classicals | | | | | | |
| 2000 $OK_{67}$ | 0.0305 ± 0.0066 | < 0.82 | 53 | > 0.16 | < 80 | -- |
| 2002 $VT_{130}$ | 0.0793 ± 0.0052 | < 0.98 | 51 | > 0.13 | < 120 | -- |
| 2001 $QD_{298}$ | 0.0487 ± 0.0059 | 1.59 ± 0.95 | 53 | $0.18\pm^{0.17}_{0.08}$ | $73\pm^{27}_{21}$ | $0.79\pm^{0.28}_{0.26}$ |
| 2001 $RZ_{143}$[d] | 0.0460 ± 0.0074 | < 0.66 | 52 | > 0.23 | < 80 | -- |
| 2001 $QS_{322}$ | < 0.0035 | < 0.97 | 52 | > 0.15 | < 100 | -- |
| Inner Classicals | | | | | | |
| 2001 $QT_{322}$ | 0.0405 ± 0.0052 | < 0.98 | 57 | > 0.21 | < 80 | -- |
| 2002 $KX_{14}$ | 0.0786 ± 0.0079 | 2.22 ± 1.44 | 54 | $0.60\pm^{0.36}_{0.23}$ | $90\pm^{25}_{19}$ | $0.61\pm^{0.28}_{0.28}$ |
| Resonant KBOs | | | | | | |
| 2001 $QR_{322}$ | 0.1684 ± 0.0099 | 3.01 ± 0.53 | 66 | $0.058\pm^{0.029}_{0.016}$ | $66\pm^{12}_{12}$ | $1.16\pm^{0.21}_{0.22}$ |
| 2003 $QX_{111}$ | 0.0189 ± 0.0070 | 4.43 ± 1.25 | 54 | $0.018\pm^{0.017}_{0.009}$ | $217\pm^{66}_{43}$ | $2.97\pm^{0.74}_{0.54}$ |

[a]Color-corrected fluxes, uncertainties, and limits. Uncertainties and limits stated are the 1σ measurement uncertainties and do not include the MIPS calibration uncertainty here.
[b]These approximate interpolated temperatures are a tool for calculating the color correction and are not to be used for thermal modeling.
[c]1σ uncertainties and limits from the STM modeling. In this paper, η encompasses physical complexities not included in the STM.
[d]Known binary objects (Noll et al. 2008a). The radii presented are effective radii such that the projected area is the same as that given by the object and its companion assuming they have equal albedos.
[e]Known presence of rotational lightcurve (Ortiz et al. 2003, Sheppard et al. 2008). The radius presented is an effective radius assuming a spherical body.

Four of the cold Classical KBOs and one of the inner Classical KBOs were not detected in the 70 μm channel. Non-detections imply that the objects' albedos are much higher than expected and radii are correspondingly smaller. As a result, the allotted exposure times were not long enough to detect the fainter than expected infrared thermal emissions. If the low inclination Classical KBOs had low albedos like the hot Classicals, then they would have been detected by *SST*.

Table 5
Monte Carlo Correction Factors for Marginal Detections

| Provisional Designation | R[a] | $p_V$[a] | $\eta$[a] |
|---|---|---|---|
| 2001 $KA_{77}$ | 1.001 | 0.9921 | 1.0112 |
| 2002 $GJ_{32}$ | 0.8933 | 1.2421 | 0.9615 |
| 1996 $TS_{66}$ | 0.9699 | 1.0459 | 0.9732 |
| 2001 $QD_{298}$ | 0.8552 | 1.3658 | 0.8620 |
| 2002 $KX_{14}$ | 0.8497 | 1.3756 | 0.7790 |
| 2003 $QX_{111}$ | 1.012 | 0.9234 | 1.0375 |

[a]Preliminary radii, geometric albedos, and $\eta$ values are multiplied by these correction factors. The products are the final values stated in Table 4.

Our data suggest that dynamically cold Classical KBOs have higher albedos than hot Classical KBOs. Results from other studies of small Classical KBOs support this conclusion (Table 6, Fig. 4; Grundy et al. 2005, Stansberry et al. 2008, Grundy et al. 2009). A new study shows that there are both red and grey inner Classical KBOs and the probability that the colors of inner Classical and cold Classical KBOs come from the same parent distribution is less than 0.1% (Romanishin et al. 2008a,b). Thus in our analysis we will consider only Classical KBOs with semi-major axes greater than 39.46 AU.

Nonparametric or rank correlation tests are appropriate to test the null hypothesis that albedos are unrelated to mean inclination since the functional form of any dependence is not known a priori. We used the Spearman rank-order test and Kendall's tau test (Press et al. 1992). These tests do not consider error bars or limits so for non-detected objects we conservatively used the values of the lower albedo limits. For all the Classical KBOs (excluding the inner Classicals) in Tables 4 and 6, the result is rejection of the null hypothesis. The Spearman rank correlation coefficient is -0.608 which implies a probability of 0.000465 of the albedo and inclination being unrelated, or a 3.50$\sigma$ confidence that they are related. The Kendall's tau is -0.471 with a significance of 0.000334 corresponding to a 3.59$\sigma$ confidence. Neither our data alone nor the data in the literature reach the 3$\sigma$ threshold but combining them allows us to conclude that albedo and inclination are indeed related among the Classical KBOs. To explore that dependence, we split the sample into two groups at an arbitrary mean inclination and applied a two-tailed Kolmogorov-Smirnov test to assess the probability that the albedo distributions of both groups are drawn from the same parent distribution. For mean inclination boundaries between 2.4° and 8.8°, that hypothesis is rejected at above the 3$\sigma$ level. This suggests that the boundary, the inclination at which an object has equal probability of belonging to the higher albedo cold Classicals or the lower albedo hot Classicals, lies within that approximate mean inclination range. Similar conclusions have been reached on the basis of colors (Gulbis et al. 2006) and binarity rates (Noll et al. 2008a); however, Peixinho et al. (2008) found a color-inclination break at about 12°. Unfortunately, our limited sample of eleven (non-inner) Classical KBOs contains only two objects with inclinations greater than 9° so we cannot determine if the albedo break differs from this color break.

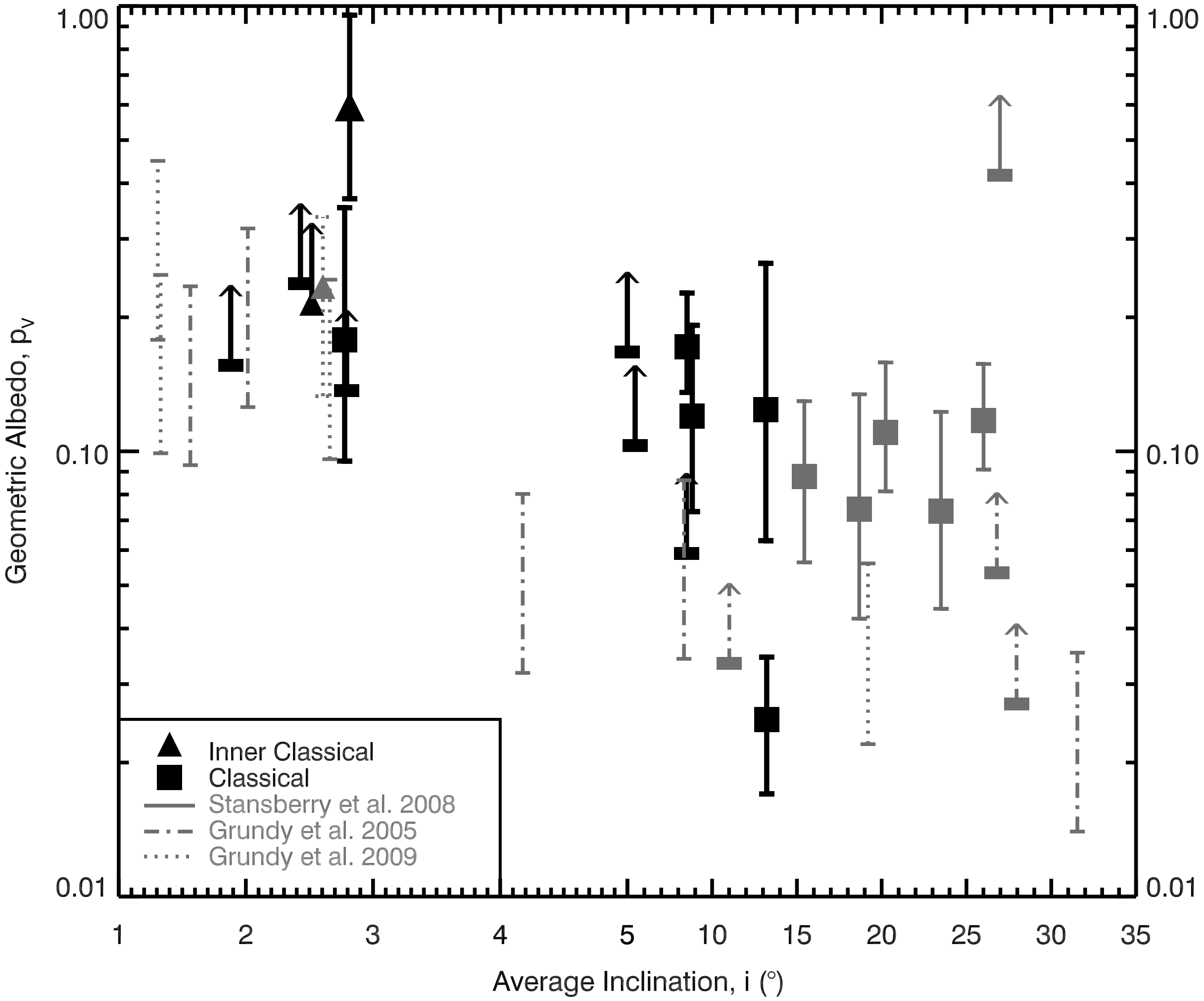

Figure 4. The visual geometric albedo, $p_V$, is plotted versus the 10 Myr mean inclination, *i*, with respect to the invariable plane. The x-axis changes scale below 5˚ to magnify the low inclination region. All included objects are Classical according to Gladman et al. (2008). Our thirteen Classical objects are plotted in black. Objects with 70 μm non-detections are represented by arrows with bars (or a triangle) located at lower limits. Comparison objects from other studies are plotted in gray. Gray objects with solid error bars are from Stansberry et al. (2008). The dot-dashed lines are constraints from Grundy et al. (2005) converted to V albedos using colors found in the literature. The dotted lines are constraints from Grundy et al. (2009). Objects with only a lower limit are represented by an arrow with a bar at the lower limit.

Classical KBOs (according to Gladman et al. 2008) are shown in Fig. 4 to illustrate that the geometric albedo is higher for Classical KBOs with lower mean inclinations. Of the thirteen Classicals in our sample, those with available V-R colors were red. Evidently, red objects may have high or low albedos. Stansberry et al. (2008) found a correlation between albedo and perihelion for their sample of Centaurs and TNOs. The objects of interest to our study are at the limit of *SST*'s sensitivity range; therefore, we could only attempt to observe the largest and closest available Classical KBOs resulting in a bias toward higher eccentricities and objects near periapsis with a strong dependence on heliocentric distance. This observational bias precludes testing for a correlation between albedo and semi-major axis, aphelion, perihelion, or object radius.

One of the objects we observed is the bright, dynamically hot KBO Quaoar. As a consistency check, the albedo and radius derived by Stansberry et al. (2008) from different *SST* observations agree with our results for Quaoar and show that physical parameters determined from *SST* observations are repeatable at moderate SNRs consistent with those attainable for KBOs. We report a smaller radius and higher albedo than that of Brown and Trujillo (2004). Brown and Trujillo used PSF modeling to resolve the size of Quaoar from images taken with the Hubble Space Telescope (HST) High Resolution Camera (HRC). They found a radius of 630 ± 95 km, a red albedo of $0.092 \pm^{0.036}_{0.023}$, and a blue albedo of $0.101 \pm^{0.039}_{0.024}$ compared to our effective radius of $454 \pm^{56}_{59}$ km and visual albedo of $0.172 \pm^{0.055}_{0.036}$. The discrepancy in radius may be related to lightcurve effects. The sixteen observations of Brown and Trujillo (2004) were taken consecutively in one HST orbit and thus did not sample the variations of Quaoar's lightcurve. Our twelve visits to Quaoar sample the period phase space well for both the double-peaked period of 17.7 h and the single-peaked period of 8.8 h (Ortiz et al. 2003) hence we were able to average over the lightcurve. Additional factors which could have contributed to different radii include the center-to-limb profile used by Brown and Trujillo and our use of the STM instead of a more realistic thermal model. Another independent size measurement such as from a stellar occultation may resolve the matter in the future.

Stansberry et al. (2008) observed the inner Classical KBO 2002 $KX_{14}$ with both the 24 μm and 70 μm channels of MIPS *SST*. However, they did not detect the object at either wavelength. They determined a lower limit on the albedo of 0.088 and an upper limit on the radius of 232 km. We were able to detect 2002 $KX_{14}$ and determined an albedo of $0.60 \pm^{0.36}_{0.23}$ and a radius of $90 \pm^{25}_{19}$ km. These results are consistent with the very loose constraints of Stansberry et al. (2008).

## 6. Discussion

Our observations imply that inner Classical and cold Classical KBOs may have much higher albedos than previously assumed (e.g. Jewitt and Luu 1993, Bernstein et al. 2004). This is evident in the plot of albedo versus inclination (Fig. 4). High albedo, red color, low absolute magnitude, and high rate of binarity are all characteristics that differentiate cold Classical KBOs from hot Classical KBOs. Cold Classical KBOs appear to be members of a homogeneous group that clearly differs from hot Classical KBOs. Inner Classicals do not share the same color distribution as cold Classical KBOs (Romanishin et al. 2008a,b) and thus even though both groups appear to have high albedos they should be considered separately.

Our experiment was not designed to rule out the presence of high albedo hot Classical KBOs. We expect that a few high albedo hot Classical KBOs may exist since that population appears to be a diverse group. Its objects have a mix of color indices and some have binary companions.

None of the cold Classical objects in our sample have large radii. These findings are consistent with the size distribution presented by Bernstein et al. (2004) in which the largest cold Classical KBO is appreciably smaller than the largest dynamically excited KBO.

Table 6
Other Small Classical KBOs with Albedo and Radius Measurements

| Number[a] | Provisional Designation | DES type[b] | $i^c$ (°) | V-R | $p_V$ | | Radius (km) | | Ref. | Name |
|---|---|---|---|---|---|---|---|---|---|---|
| **Hot Classicals** | | | | | | | | | | |
| 55636 | 2002 $TX_{300}$ | SN | 26.98 | 0.363±0.018[i,j,k] | > 0.41 | | < 210 | | f | |
| 55565 | 2002 $AW_{197}$ | SN | 26.01 | 0.62±0.03[l] | $0.115 \pm^{0.041}_{0.025}$ | | $371 \pm^{49}_{52}$ | | f | |
| 90568 | 2004 $GV_9$ | SN | 23.49 | 0.52±0.03[m] | $0.073 \pm^{0.049}_{0.029}$ | | $342 \pm^{34}_{37}$ | | f | |
| 55637 | 2002 $UX_{25}$[d] | SN | 20.23 | 0.56±0.02[m] | $0.111 \pm^{0.049}_{0.030}$ | | $340 \pm^{53}_{54}$ | | f | |
| | 2002 $MS_4$ | SN | 18.68 | 0.38±0.02[m] | $0.073 \pm^{0.058}_{0.032}$ | | $365 \pm^{59}_{60}$ | | f | |
| 20000 | 2000 $WR_{106}$ | C | 15.44 | 0.615±0.014[n] | $0.088 \pm^{0.042}_{0.031}$ | | $357 \pm^{89}_{64}$ | | f | Varuna |
| 50000 | 2002 $LM_{60}$[d] | C | 8.52 | 0.646±0.012[k,l] | $0.206 \pm^{0.093}_{0.066}$ | | $415 \pm^{89}_{71}$ | | f | Quaoar |
| **Inner Classicals** | | | | | | | | | | |
| 119951 | 2002 $KX_{14}$ | C | 2.82 | 0.621±0.022[o,p] | > 0.10 | | < 220 | | f | |
| | | | | | Min. | Max. | Min.[e] | Max.[e] | | |
| **Hot Classicals** | | | | | | | | | | |
| | 2001 $QC_{298}$[d] | SN | 31.54 | -- | 0.014 | 0.035 | 94 | 150 | g | |
| 19308 | 1996 $TO_{66}$ | SN | 27.95 | 0.397±0.028[q,r,s,t] | 0.027 | 1 | 0 | 449 | g | |
| 24835 | 1995 $SM_{55}$ | SN | 26.79 | 0.395±0.026[n,r,s,u] | 0.053 | 1 | 0 | 350 | g | |
| | 2004 $PB_{108}$[d] | SN (UN) | 19.19 | -- | 0.02 | 0.06 | 99 | 160 | h | |
| 19521 | 1998 $WH_{24}$ | C | 11.01 | 0.623±0.032[n,r,u,v] | 0.033 | 1 | 0 | 372 | g | Chaos |
| | 1998 $WW_{31}$[d] | C | 8.34 | -- | 0.034 | 0.086 | 59 | 93 | g | |
| **Cold Classicals** | | | | | | | | | | |
| 88611 | 2001 $QT_{297}$[d] | C | 4.18 | 0.61±0.04[w] | 0.032 | 0.080 | 65 | 103 | g | Teharonhiawako |
| | 2001 $XR_{254}$[d] | C | 2.66 | -- | 0.10 | 0.24 | 68 | 110 | h | |
| 58534 | 1997 $CQ_{29}$[d] | C | 2.02 | 0.667±0.094[r,s,t,x] | 0.126 | 0.317 | 30 | 47 | g | Logos |
| 66652 | 1999 $RZ_{253}$[d] | C | 1.56 | 0.688±0.094[u] | 0.093 | 0.235 | 66 | 104 | g | Borasisi |
| 134860 | 2000 $OJ_{67}$[d] | C | 1.33 | 0.673±0.046[n] | 0.10 | 0.25 | 57 | 90 | h | |
| | 2003 $TJ_{58}$[d] | C (UN) | 1.31 | -- | 0.18 | 0.45 | 26 | 42 | h | |
| **Inner Classicals** | | | | | | | | | | |
| | 1999 $OJ_4$[d] | C | 2.61 | 0.668±0.072[y] | 0.13 | 0.34 | 29 | 47 | h | |

[a]Objects are sorted by Gladman et al. (2008) orbital type and decreasing inclination with a break at 5° to separate the overlapping hot and cold Classical KBO populations. Makemake and Haumea were not included as they are known to be larger and distinct.
[b]SN – Scattered-Near, C – Classical, UN – Unclassified in Gladman et al. (2008).
[c]Average inclinations are with respect to the invariable plane.
[d]Known binary objects (Noll et al. 2008a).
[e]For binary objects, the minimum and maximum radius stated are for the brighter component assuming that the primary and secondary have equal albedos.
[f]Albedo and radius results from Stansberry et al. (2008) have been remodeled from their Spitzer flux measurements using our method.
[g]Albedos and radii from Grundy et al. (2005). If the object is designated binary, then its albedo and radius were determined from the binary system mass. If the object is not designated binary, then its albedo and radius were determined from thermal radiometry.
[h]Albedos and radii from binary system mass (Grundy et al. 2009).
[i]Doressoundiram et al. (2005).
[j]Ortiz et al. (2004).
[k]Tegler et al. (2003).
[l]Fornasier et al. (2004).
[m]Tegler et al. http://www.physics.nau.edu/~tegler/research/survey.htm
[n]Doressoundiram et al. (2002).
[o]The object 2002 $KX_{14}$ was observed in the V and R bands by S. Sheppard using filters based on the Johnson system with four 300 second images in each filter at the du Pont 2.5 meter telescope on UT July 19, 2007 with the Tek5 CCD (0.259" $pixel^{-1}$).
[p]Romanishin et al. (2008b).
[q]Barucci et al. (1999).
[r]Boehnhardt et al. (2001).
[s]Gil-Hutton and Licandro (2001).
[t]Jewitt and Luu (2001).
[u]Delsanti et al. (2001).
[v]Barucci et al. (2000).
[w]Osip et al. (2003).
[x]Stephens et al. (2003).
[y]Hainaut and Delsanti (2002).

The mass distribution of the cold Kuiper Belt should be revised in light of our new high albedo results. A few years ago, KBOs were assumed to have geometric albedos equal to 0.04 unless measurements had been taken. This underestimation in albedo led to an overestimation in radius which in turn led to a significant overestimation in mass. Stansberry et al. (2008) found an average Spitzer-derived albedo for KBOs of 0.0988 for a sample that does not contain any low inclination non-Resonant KBO detections. Here we find that cold Classical KBOs have higher albedos than those found by Stansberry et al. (2008). To illustrate the effect on estimated masses, consider a cold Classical KBO with an actual geometric albedo of 0.20 assumed to be 0.10 as for the KBOs in Stansberry et al. (2008). The albedo has been scaled by a factor of $2^{-1}$ (underestimated). If the canonical value of the phase integral ($q$=0.39) is used instead of our linear fit ($q = 0.336 * p_V + 0.479$), then the radius as calculated via the Bond albedo has been scaled by a factor of $2^{1/2}$ (overestimated) and the mass has been scaled by a factor of $2^{3/2}$ (overestimated). If a KBO has an albedo of 0.40 instead of 0.10, then the error would be even greater: the albedo has been scaled by a factor $4^{-1}$ (underestimated), the radius scaled by a factor of 2 (overestimated), and the mass scaled by a factor of 8 (overestimated). Applying a higher albedo to the results of Bernstein et al. (2004), the dynamically cold KBO population would have a mass of approximately $8*10^{-4}$ $M_\oplus$ with an albedo of 0.20 as opposed to $9*10^{-3}$ $M_\oplus$ with an albedo of 0.04 (a reduction by a factor of $5^{-3/2}$ or approximately 0.1). If we consider that some TNO binaries have densities less than 1 g $cm^{-3}$ (Noll et al. 2008a), then it is possible the total mass of the cold Classical KBO population could be even less.

The Kuiper Belt is the accepted source of Jupiter Family Comets (JFCs) which have very small radii and albedos of 0.04 on average. Duncan et al. (2004) and references therein showed that the main source region for JFCs is the scattered disk. Using the magnitude distribution of Bernstein et al. (2004), Volk and Malhotra (2008) have shown that there is a discrepancy between the observational estimate of comet-sized objects in the scattered disk population and the number of objects necessary to supply the JFCs in steady state. The Classical Kuiper Belt may be a supplementary source of JFCs. The high albedos and red colors of cold Classical KBOs differ from those of JFCs (Jewitt 2002) but these are not a major concern. The albedos and radii of Classical KBOs may be reduced through collisional fragmentation resulting in structural alteration (Farinella and Davis 1996, Davis and Farinella 1997, Pan and Sari 2005). If fragments of cold Classical KBOs do find their way into the inner Solar System, then presumably their albedos decline and colors become less red with exposure to more intense sunlight, possibly through cometary activity (Jewitt 2002, Delsanti et al. 2004). This albedo trend is consistent with the general decline in albedo from the Kuiper Belt through the Centaurs and on to the JFCs (Stansberry et al. 2008, Grundy 2008).

## 7. Conclusions

We have found that the inner Classical and cold Classical KBOs in our sample have high geometric albedos. For the sample of main belt Classical KBOs listed in Tables 4 and 6, albedo is highly correlated to inclination at the $3.5\sigma$ level. This strengthens the hypothesis that dynamically cold and hot KBOs have different origins and/or evolutionary histories since we did not find consistently high albedos among our hot Classical KBOs. Different albedo distributions add to the already strong evidence from orbital dynamics, color photometry, and rates of binarity. Our combined sample of hot and cold Classical KBOs contained mostly red objects thus red Classical KBOs have a range of albedos.

The previous underestimation in albedo has led to an overestimation of the total mass of the cold Classical KBO population by an order of magnitude. Future investigations into the dynamical evolution of the Kuiper Belt and its mass distribution and size distribution will need to account for the reduction in mass implied by high albedos.

It may be more appropriate to use a linear function for the phase integral rather than adopting the canonical asteroid value of 0.39, especially for cold Classical KBOs with high albedos. A functional form of the phase integral will more accurately convert geometric albedos to Bond albedos which are used to construct temperature distributions in thermophysical modeling. If the Bond albedo is underestimated by using the canonical value for the phase integral, then the surface temperature and thermal emissions are overestimated.

It is imperative for further study of TNOs to know their sizes and albedos. Accurate albedo and radius derivations allow scientists to interpret photometric color and spectral observations unambiguously, to compute rigorous size frequency distributions, to determine the densities of objects whose masses have been determined from the orbits of satellites, and to compute escape velocities, surface temperatures, and volatile loss rates. Our work increases the number of small Classical KBOs with accurate sizes and albedos from 21 to 32. We look forward to the results of future observations from *SST*, the Herschel Space Observatory, the James Webb Space Telescope (JWST), and the New Horizons mission. In particular, if New Horizons flies by a TNO after its flyby of Pluto, then that TNO will most likely be a cold

Classical KBO allowing measurement of its radius, albedo, and phase integral. Knowing that cold Classical KBOs have high albedos will be invaluable to planning the encounter.

**Acknowledgments**
This work is based on observations made in Cycle 1 program P3542 with the Spitzer Space Telescope, which is operated by the Jet Propulsion Laboratory, California Institute of Technology under a contract with NASA. Support for this work was provided by NASA through an award issued by JPL/Caltech. We are grateful to Lowell Observatory for instituting their pre-doctoral fellowship program. We thank L. Wasserman, R. Millis, and D. Cruikshank for their input. We also wish to thank the reviewers,Amanda Gulbis and an anonymous reviewer, for their comments and suggestions.